# Is Rust Used Safely by Software Developers?

Ana Nora Evans
AnaNEvans@virginia.edu
University of Virginia

Bradford Campbell
bradjc@virginia.edu
University of Virginia

Mary Lou Soffa
soffa@virginia.edu
University of Virginia

## Abstract

Rust, an emerging programming language with explosive growth, provides a robust type system that enables programmers to write memory-safe and data-race free code. To allow access to a machine's hardware and to support low-level performance optimizations, a second language, *Unsafe Rust*, is embedded in Rust. It contains support for operations that are difficult to statically check, such as C-style pointers for access to arbitrary memory locations and mutable global variables. When a program uses these features, the compiler is unable to statically guarantee the safety properties Rust promotes. In this work, we perform a large-scale empirical study to explore how software developers are using *Unsafe Rust* in real-world Rust libraries and applications. Our results indicate that software engineers use the keyword **unsafe** in less than 30% of Rust libraries, but more than half cannot be entirely statically checked by the Rust compiler because of *Unsafe Rust* hidden somewhere in a library's call chain. We conclude that although the use of the keyword **unsafe** is limited, the propagation of unsafeness offers a challenge to the claim of Rust as a memory-safe language. Furthermore, we recommend changes to the Rust compiler and to the central Rust repository's interface to help Rust software developers be aware of when their Rust code is unsafe.

## 1 Introduction

Programming languages directly impact the reliability, safety, and correctness of software, and their features impact the prevalence of bugs in actual software. A relatively new programming language, Rust, is explicitly designed to help programmers write more reliable software by using the compiler to help reduce memory and data race errors. Rust is referred to as a "safe" systems programming language, indicating that its type system, ownership model, automatic memory management without garbage collection, and static compiler make it well suited for writing lower-level or core software without the common bugs that can plague code written in C and C++ [15, 29, 30].

The design aspects that make Rust safe, such as no arbitrary pointers or arbitrary type casting, however, would also make writing most or all low-level code impossible. Operations such as configuring hardware or reading a network socket involve manipulating memory in ways that the compiler cannot guarantee to be safe.

Therefore, Rust includes an "escape hatch" with the **unsafe** keyword[1] that allows programmers to deactivate some (but not all) of the Rust compiler's checks for certain regions of code. This functionality was originally described as "pragmatic safety" [14] when Rust was first introduced, and allows developers to use their own discretion when writing Rust code. Part of the justification for allowing *Unsafe Rust* code is that uses of **unsafe** would be easy to locate and audit, and that developers can decide how much unchecked code they are willing to accept in their software.

Modern software development leverages and builds upon libraries, which often use yet other libraries. Auditing software for uses of **unsafe** requires auditing all dependent libraries, a potentially cumbersome task. This overhead is mitigated, however, if usage of **unsafe** is scarce and easy to locate in the Rust software ecosystem, or if making a determination about the validity of the **unsafe** usage is typically straightforward. Therefore, understanding how developers are actually using **unsafe** is necessary to evaluate whether "pragmatic safety" is valid and if Rust provides a safe programming environment in practice.

Our study is further motivated by recent interest in Rust as a safe alternative to C for systems software [3, 6, 8, 20, 21, 24, 33] and by the development of formal definitions for Rust's type system (including *Unsafe Rust*). For example, the Rust Belt project [18] proposes formal tools for verifying *Unsafe Rust*, and Oxide [39] presents a formalization of a language very similar to Rust. The Rust open source community recently formed a new Rust working group to create a "Unsafe Code Guideline Reference" to help guide developers [36]. These are encouraging steps, and understanding how *Unsafe Rust* is being used by developers will help guide the successful formation of these strategies.

To acquire this understanding, we perform a large-scale study and an analysis of publicly available Rust libraries and application code. We first determine how frequently the **unsafe** keyword is used. Then we analyze the call graph of every function in our data set to identify if at any point the function may use code that is not safe and not checked by the compiler. This analysis enables us to find code that *looks* safe, but is actually *Unsafe Rust*. To better understand how developers are using **unsafe**, we also identify the underlying code behavior that necessitates the use of *Unsafe Rust* to analyze the frequency of the various **unsafe** operations. Further, we observe the use of *Unsafe Rust* over time to see if there are evolving changes in the community.

To perform this analysis, we developed and implemented an algorithm for constructing an extended call graph of Rust functions that uses the type information to increase the call graph precision. Building a call graph for Rust is difficult, however, as Rust's runtime polymorphism and higher order functions complicate statically building a call graph without missing edges or adding extra edges. In our approach, we identify these ambiguities, and build two versions of the call graph by applying both a conservative (assume the call will be to unsafe code) and an optimistic (assume the call we be to code statically checked to be safe) analysis to help bound the use of *Unsafe Rust* in Rust libraries. We then traverse and analyze the resulting extended call graph to determine how unsafeness propagates in real-world Rust code.

---

[1]Rust keywords are green and bolded while the safe and unsafe conditions are italicized.

After analyzing over 85% of the valid Rust libraries available at the start of our study, we find that 29% contain at least one explicit use of **unsafe**. When considering the dependency tree, however, that number increases to around 50%, meaning half of Rust libraries use *Unsafe Rust* or rely on other libraries that use *Unsafe Rust*. Narrowing down to just the most used and downloaded libraries increases the use of *Unsafe Rust*, as around 60% of popular crates include *Unsafe Rust*. The majority of **unsafe** uses in the Rust ecosystem are to call other Rust functions that are marked **unsafe**. We find that only 22% of these **unsafe** functions are to external libraries implemented in C, suggesting that a majority of the *Unsafe Rust* is actually from Rust code where the software developer decided to disable the compiler checks. Finally, we see negligible increases in the frequency of **unsafe** used over the past ten months.

Based on these findings, we propose several recommendations to help realize the original vision of "pragmatic safety" in Rust, including programmer-assisted automated checks, additional tooling to help developers identify uses of **unsafe**, and more visible code reviews to audit uses of **unsafe**. These changes would allow developers to better exploit the benefits of *Unsafe Rust* while managing its risks to overall software reliability.

The contributions of this paper are:

- A large-scale study and analysis of **unsafe** use in the Rust software ecosystem.
- A call graph construction algorithm which handles generic polymorphism.
- Findings that indicate that a majority of crates are not guaranteed by the compiler be memory-safe and data-race free.
- Highlights of our findings indicate that unsafe function calls are the most common use of unsafeness, which is caused through library dependencies rather than the use of the unsafe keyword, and the most downloaded crates have more unsafe code than other other crates;
- Recommendations to enhance the Rust compiler and its ecosystem to help developers understand when they are using *Unsafe Rust* in their software.

Section 2 contains a brief introduction to Rust, *Unsafe Rust* (see Section 2.3), the Rust software ecosystem (see Section 2.6), and the terms we define to facilitate discussion throughout of paper (see Section 2.5). In Section 3 we present the research questions that guided our experiment and the selection criteria for the Rust code that comprises our data set. Section 4 contains the approach to answering the research questions. In Section 5 we define the data set we used in the experiments, and the answers to the research questions based on the experimental results. We conclude with our recommendations (see Section 6), threats to validity (see Section 7), related work (see Section 8) and conclusions (see Section 9).

## 2 Background

First, we provide a working example that will be referenced throughout the paper. We then describe the key features of *Safe Rust* and *Unsafe Rust* that are relevant to our study. Next, we define terms that we use to categorize Rust code. Finally, we provide some background information about the general Rust software ecosystem.

### 2.1 Working Example

Figure 1(a) shows a set of functions in pseudocode from several libraries selected to illustrate how unsafety can propagate in a codebase. We will use these functions to explain concepts throughout the paper.

The functions are organized in five different libraries. The starting point is function foo() in Library1, which calls the function bar() in Library2. The symbol :: separates the library name from the function called from that library. In function bar() from Library2, the call to my_object.baz() is a run-time polymorphic call to the method baz() of an instance implementing the interface HasBaz. The interface HasBaz has two implementations in Library3 and Library5.

Figure 1(b) shows the resulting library dependency graph from the example functions. A library depends on other libraries if it uses functions from those libraries, and therefore requires the other libraries to completely compile a binary.

For functions with run-time polymorphism we generate two call graphs: conservative and optimistic (see Figure 1(b)). We will explain this further in Section 4.

### 2.2 Safe Rust

Rust includes a few basic concepts that enable the compiler to enforce safety guarantees. The *ownership* mechanism in *Safe Rust* requires that a unique variable is the owner for every memory location. Memory locations are immutable unless explicitly declared otherwise. Variable assignment results in a copy or a transfer of ownership, and once the variable loses ownership of the memory location, that variable becomes unusable. To enable sharing, the *borrow* mechanism allows creating memory aliases which permit any number of read-only references and exactly one mutable one. The foo() function in Figure 1(a) shows an example borrow operation. Together, the *ownership* and *borrow* mechanisms prevent a large class of memory-safety errors and data races.

The definition of memory-safety used by Rust is similar to the one proposed by Szekeres *et al.* [30]. A Rust program is memory-safe if it is free of any memory errors such as dereferencing a null or dangling pointer, reading or writing unaligned pointers, and reading uninitialized memory [34]. Memory leaks are explicitly considered defined behavior, and thus not memory errors.

For instances where program operation requires violating these constraints, *Unsafe Rust* allows developers to assert to the compiler that they are manually implementing the necessary checks to preserve memory-safety and data-race freedom.

### 2.3 Unsafe Rust

*Unsafe Rust* provides the necessary operations for low-level systems programming, such as arbitrary memory accesses with C-style pointers, invoking system calls, calling foreign functions (usually C functions), executing inline assembly instructions, eliding bounds checks for performance, and accessing global static memory. To use any of these features, developers must mark the code with the **unsafe** keyword. In Listing 1, **unsafe** is necessary for casting an address to a Rust struct for memory-mapped IO, and to use global variables to reference unique resources (e.g., COM1 port).

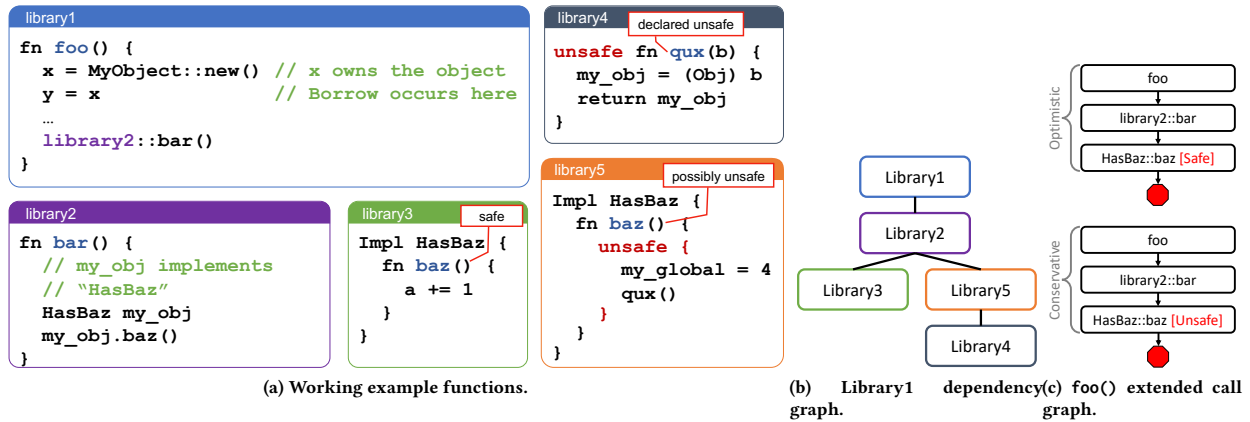

Figure 1: Working example of Rust functions in multiple libraries that are either safe, unsafe, or only appear safe.

```
1  impl SerialPort {
2      pub unsafe fn new(base: usize)
3              -> &'static mut SerialPort {
4          &mut *(base as *mut Self)
5      }
6  }
7  pub static COM1: Mutex<SerialPort> =
8      Mutex::new(SerialPort::new(0x3F8));
9  pub unsafe fn init() {
10     COM1.lock().init();
11 }
```

Listing 1: Transmute

Within an unsafe region, the compiler still checks the *Safe Rust* types, but the operations listed above are permitted as well. *Unsafe Rust* does not grant the programmer complete freedom, but subverting overall system safety is certainly possible. Programmers using *Unsafe Rust* are responsible for writing code free of safety violations and undefined behavior; however, what constitutes undefined behavior is currently not well specified and can change with different versions of the compiler. This situation makes safely using *Unsafe Rust* difficult.

### 2.4 Sources of Unsafe Rust

There are several sources of *Unsafe Rust*, including unsafe operations, unsafe functions, and unsafe traits.

**2.4.1 Unsafe Operations** A developer may directly use *Unsafe Rust* by creating a code block labeled with the keyword **unsafe**, which is required for the following operations:

(1) Calling a function marked **unsafe**, non-Rust external function, or a compiler intrinsic (a function whose implementation is handled specially by the compiler).
(2) Dereferencing a C-style pointer.
(3) Accessing a mutable global variable.
(4) Using inline assembly instructions.

(5) Accessing a field of a union type.

Function baz() in Library5 contains an example of *Unsafe Rust* operation: an assignment to a global variable my_global, enclosed in an **unsafe** block.

An example of a possibly dangerous **unsafe** function call is the mem::transmute() function used to coerce the contents of an arbitrary memory location into a specific Rust type. This is necessary when raw data (such as from a network socket), but can easily violate type-safety if used improperly. Further, the mem::transmute function also makes use-after-free memory errors possible when it is used to extend the compiler calculated code bounds where the variable is live.

```
1  let mut hello = String::new();
2  let hello_ref: &mut String =
3      {let r: *mut String = &mut hello;
4      unsafe { &mut *r }}; //Undefined Behavior! A second mutable reference
5  hello.push_str("Hello ");
6  hello_ref.push_str("world!");
7  println!("{:?}", hello);
```

Listing 2: Multiple Mutable References

Using **unsafe** also makes mutable reference aliasing possible, leading to undefined behavior. In the Rust Listing 2, using simplified Rust syntax, two mutable references (hello and ref1) are created to the same memory location using an **unsafe** block. Different Rust versions are free to handle this differently, and do, as the Rust compiler version 1.36.0 and the Rust Mid-Level Intermediate Representation Interpreter (Miri) version 0.1.0 handle code with multiple mutable references to the same string differently. If the code is compiled and executed, the output the programmer may expect ("Hello world!") is printed. However, when interpreted by Miri an error is displayed, as Miri also includes numerous checks of memory safety and that the aliasing rules for references are not violated [17]. This suggests extensive care is required to use **unsafe** with maintainable and reliable code.

If a block of code is marked **unsafe**, but does not actually contain any unsafe operations, the the compiler emits a warning message. We assume that the Rust developers remove necessary unsafe block labels as a result.

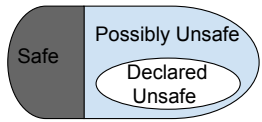

Figure 2: Function annotation taxonomy.

#### 2.4.2 Unsafe Functions
A Rust function definition may be labeled with the **unsafe** keyword (for example the qux() function in Figure 1(a)) to signal to a caller that preconditions must be satisfied or memory-safety errors may occur. Thus, an **unsafe** function may contain only *Safe Rust* operations but still be marked unsafe and the compiler will not emit an warning because it is unable to distinguish between programmer's intent and a mistake. For example, a function that sets the length of a vector, Vector.set_length(), will be marked **unsafe** despite only setting an internal field because if it is called with a parameter larger than the capacity of the internal buffer, future vector operations may access unallocated memory. This example also demonstrates how memory safety errors may occur in *Safe Rust* code after any use of *Unsafe Rust*.

Conversely, a function containing **unsafe** operations may provide a safe interface. For example, array operations are commonly implemented using C-style pointers and include their own safety checks. In these cases Rust developers are expected to analyze the code manually to ensure safety is preserved.

#### 2.4.3 Unsafe Traits
A Rust **trait** [34] describes an abstract interface that types may implement. A trait may be declared **unsafe** if it contains any **unsafe** method or its implementations must satisfy an invariant. An example from the Rust standard library is the trait Send, and any type that implements Send declares that it is safe to change the ownership of the type to another thread. An implementation of an **unsafe** trait, must be marked **unsafe**. The programmer must ensure that the implementation satisfies the trait's invariant, as the compiler cannot automatically check this.

### 2.5 Function Taxonomy
To explain our analysis procedure and results, we clarify terms that we use to label Rust code with regards to its safety of functions.
**Declared unsafe:** A Rust function whose declaration includes the **unsafe** keyword is *declared unsafe*. In Figure 1(a) the qux() function in Library4 is explicitly declared unsafe. A call to such a function must be in an **unsafe** block or another declared unsafe function. In our working example, the function baz() in Library5 calls qux() from an **unsafe** block.
**Possibly unsafe:** A Rust function that may execute *Unsafe Rust*. A function may be *possibly unsafe* if: (1) it is *declared unsafe*; (2) it contains an **unsafe** block; or (3) it calls a *possibly unsafe* function. In Figure 1(a), the foo() function in Library1 is possibly unsafe because it may use the baz() implementation from Library5.
**Safe:** A Rust function with compile-time guarantees of memory-safety and data-race freedom is a safe function. The baz() function in Library3 of Figure 1(a) is safe.

The relationship among the different sets is presented in Figure 2. All Rust functions are either safe or possibly unsafe, and all declared unsafe functions are possibly unsafe.

### 2.6 Rust Software Ecosystem
*Crates* are the unit of compilation in Rust, meaning they compile individually and the build output may be a software library or an binary application, depending on if the crate includes a main() function. Crate compilation is managed by Rust's build and package manager *cargo*. Crates typically rely on cargo, making the build process standard in the Rust ecosystem and easing large-scale automated analysis. Publicly available Rust crates, either libraries or applications, are typically published in crates.io, a central repository of Rust software.

## 3 Research Methodology
In this section, we present the research questions that guide our study and the selection criteria for the Rust code we explore.

### 3.1 Research Questions
The main goal of our study is to gain insights on the practical use of *Unsafe Rust* by software engineers and the consequences for the static safety guarantees of typical Rust code. The Rust compiler generates some code with unsafe operations (*e.g.*, the PartialEq trait implementation of enumerations). Our analyses do not include compiler generated unsafe and consider only code written by a developer.

**RQ1:** How much do developers use *Unsafe Rust*?
We count the uses of the **unsafe** keyword for each of the four abstractions (block, function, trait, and trait implementation). The motivation of RQ1 is to understand how often Rust developers explicitly use **unsafe** in their code, and, thus, lose the static safety guarantees.

**RQ2:** How much of the Rust code is *Unsafe Rust*?
*Unsafe Rust* is necessary, but undermines the memory-safety and data-race freedom guarantees. Since *Unsafe Rust* may modify an arbitrary memory location, any use of *Unsafe Rust* in any dependency compromises the static guarantees of the whole library. Rust libraries that do not contain *Unsafe Rust* operations, but call functions from external libraries that do, appear *Safe Rust*, but, in fact, are not statically guaranteed safe. The goal of RQ2 is to determine what percentage of the Rust libraries lack full safety compiler guarantees.

**RQ3:** What *Unsafe Rust* operations are used in practice?
We locate and count the different types of *Unsafe Rust* operations in **unsafe** blocks and *declared unsafe* functions separately. The motivation for RQ3 is to understand if interactions with C are the main source of *Unsafe Rust*. If yes, then most *Unsafe Rust* can be eliminated by implementing the libraries in Rust. If not, then the reason for *Unsafe Rust* lies within Rust code itself, and may be necessary for achieving the desired performance or for implementing low-level libraries that interact with the operating system or directly with hardware.

**RQ4:** What abstract binary interfaces (programming languages) are used in the *declared unsafe* functions?
The motivation for RQ4 is to understand if most called **unsafe** functions are from external libraries implemented in C or from other Rust libraries.

**RQ5:** Does the use of *Unsafe Rust* change over time?
The motivation for RQ5 is to understand if the increased attention in the past year from the research community and the efforts by Rust language team to define and develop guidelines for use of **unsafe** changed the habits of the Rust developers.

**RQ6:** Why do Rust developers use **unsafe**?
We surveyed Rust developers to understand their reasons for using **unsafe**.

### 3.2 Data Selection

To understand unsafety in Rust, we analyze real-world, publicly available Rust code. As the Rust tool chain provides robust support for libraries, Rust software extensively leverages community-provided libraries to create larger projects. As such, we target our analysis towards the libraries that comprise the Rust software ecosystem.

To provide as wide of an analysis as possible, we include as many libraries (Rust crates) in our study as possible. However, as with any open ecosystem, there exists a "long-tail" of crates in Rust that are small, largely unused projects, and these may not be representative of the ecosystem at large. Therefore, we also perform our analysis on only the "popular" crates in our data set as defined as having the most downloads from the central Rust repository to identify any differences between the entire data set and the subset that is more frequently used.

Additionally, to compare crates contributed by the larger Rust community with those developed by members of the core Rust development team and Mozilla Research [32], we analyze the application Servo [3], a web browser engine from Mozilla. Servo, endemic of the larger Rust ecosystem, itself is implemented as a collection of about fifty discrete crates, and together with all its dependencies, compiling Servo involves compiling almost 400 different crates.

We include all Servo crates and their external dependencies in our analysis of Servo because they are implicitly vetted by the Servo team to be included in one of the flagship Rust applications.

## 4 Approach

Our approach to answer the research questions is to identify all occurrences of **unsafe** in Rust codebases, and then determine how the unsafeness propagates to caller functions. First, we parse Rust code from libraries to identify the keyword **unsafe** in blocks, functions, traits, and trait implementations in a single crate, which can be analyzed independently of other crates. Next, we develop two versions of an extended call graph for each library; one that is optimistic and one that is conservative in terms of whether a polymorphic function is safe or unsafe. Finally, we develop an algorithm to analyze the extended call graph to propagate the unsafe condition through the call graph to determine if a function is safe or possibly unsafe. In this way, we can identify libraries that appear safe but actually have potentially unsafe conditions that have propagated up the call chain. Sources of imprecision in our analysis include potential inclusion of dead code.

### 4.1 Extended Call Graph

In general, a program call graph [28] is a directed graph with a node for each program function and an edge $(f, g)$ for one or more potential calls of $g$ from the function $f$. This simple construction needs to be expanded if the language has polymorphism and high order functions, as Rust does. Several algorithms for call graph construction for other languages which have these features have been proposed with different trade-offs between precision and running time [5, 11, 25]. Our approach is similar to the one by Petrashko *et al.* [25] in using the type information available at call site of not only the Self type, but also of the static types of the parameters. Each node of the call graph is *extended* to contain not only the function, but a list of generic type parameters and type substitutions for those when available.

To generate a call graph, we use the Rust compiler to compile the crate to an intermediate representation of Rust known as MIR (Middle Intermediate Representation). We then use the "control flow graphs" of functions obtained directly from the MIR representation to run a context-sensitive analysis which uses type inference to find the precise functions that can be called. This analysis starts at the leaf terminal nodes of the control flow graph of a function. For each terminal leaf node of the type "function call" we use the type inference to determine the function call, given the type substitutions in the calling context. If the method has no Self type or has a Self (this) type that it is statically known, then only one edge is introduced in the call graph in the actual implementation. If a node with the calculated substitutions for generics of the called function does not exist in the call graph, then we apply the substitutions recursively and introduce new nodes in the call graph as needed. If the Self type (*i.e.*, this) is still unknown, then an edge to a node parameterized by the Self type is introduced. In building the call graph, when we encounter functions with generic type parameters, we parameterize the call graph based on the generic type parameters and instantiate a call with actual types available at the call site. The nodes in our call graph are functions together with a set of type substitutions of the generic type parameters when statically known.

If the precise functions of the call graph cannot be resolved at compile-time (due to virtual dispatch and higher order functions), we split into two approximations, and create two versions of the call graph. We create a call graph of a conservative approximation assuming that the unknown function is unsafe, and an optimistic approximation of a call graph that assumes the potential function is safe.

For efficiency, our call graph construction for a function terminates when any **unsafe** usage is found, as the original function must now be marked *possibly unsafe* and further calls will have no impact. Similarly, we stop when we reach a virtual call or a function pointer as our two approximations cover the two possible cases.

We generate extended call graphs for every function in every crate, enabling this approach to work even for libraries which do not contain a main function. These extended call graphs are for individual libraries, and function calls to other libraries are explicitly marked. In our analysis, we can then combine the extended call graphs as needed without having to generate the call graph multiple times. Note, we stop our call graph construction at calls into the Rust standard library. If the functions are *declared unsafe* we mark them as **unsafe**; however, we consider the standard library trusted and consider all other functions *safe*, even if their implementations contain uses of **unsafe**.

To aid future reproducibility studies and extensions of our work, we publish all the code used for this paper at https://github.com/

ananevans/icse2020, including the call graph construction available in the unsafe-analysis/src/implicit_unsafe/rta.rs folder.

### 4.2 Analysis on the Call Graph

---
**Data:** call graph
**Result:** list of possibly unsafe functions
**for** *all function definitions* **do**
    **if** *function has unsafe in body* **then**
        add function to possibly unsafe list;
        add function's call graph node to worklist
    **end**
**end**
reverse the call graph;
**while** *worklist not empty* **do**
    current_func = pop the first element of the worklist **for**
    *each neighbor of current_func* **do**
        **if** *neighbor not in possibly unsafe list* **then**
            add function to possibly unsafe list;
            add neighbor to worklist
        **end**
    **end**
**end**

**Algorithm 1:** Analysis on Call Graph

---

After the extended call graphs are constructed, we use them to identify *potentially unsafe* functions in the Rust libraries. Our approach starts at the bottom of the call graph with all functions that contain **unsafe** blocks. We then work back up the call graph to identify functions that may call any of these initial functions. Our goal is to mark every function as either *safe* or *possibly unsafe*.

We start with a worklist initialized with all functions that contain an **unsafe** block in their body. Note that we do not need to consider the *declared unsafe* functions because they will necessarily be called from within an **unsafe** block in another function. We iteratively remove a function from the worklist, and perform a reverse propagation on the extended call graph to mark all of its callers as *possibly unsafe*, and then add them to the worklist. The algorithm completes when the worklist is empty, and all functions that have not been marked *possibly unsafe* are deemed *safe*.

### 4.3 Example

Consider our working example from Figure 1 and assume we are analyzing the safety of Library1. The library dependency graph is given in Figure 1(b) and shows the order of compilation and analysis in building the extended call graph. Function foo() in Library1 calls a function bar() in Library2 and thus an edge is placed between Library1 and Library2. Library2 specifies libraries 3 and 5 as dependencies, making the call to polymorphic function baz() ambiguous. Finally, Library5 contains a call to qux() and therefore Library4 is a dependency of Library5.

The extended call graph with external libraries merged is shown in Figure 1(c). Because bar() calls the polymorphic function baz(), we cannot precisely determine the exact call graph and split into an optimistic call graph that assumes the unknown implementation of HasBaz::baz() is in fact safe, and a conservative call graph that assumes it is unsafe. To determine unsafety in the conservative case, we start with the unknown function HasBaz::bar() in our worklist since it is marked *possibly unsafe*. We then find all callers, in this case library2::bar(), mark them as *possibly unsafe*, and add them to the worklist. Iterating, we determine foo() calls bar() and mark foo() *possibly unsafe*. This tells us that even though foo() *appears* safe, in a conservative analysis it may in fact call code which is not statically checked by the compiler.

## 5 Experiments and Results

We start with describing the experimental setup in Section 5.1 and continue with the detailed description of the data sets used in our experiments (Section 5.2). We conclude with the results in Section 5.3.

### 5.1 Experimental Setup

We execute the experiments using version "nightly-2018-09-11" of the Rust compiler on Ubuntu 18.04. The 2019 version of the most downloaded data set is compiled with version "nightly-2019-07-01".

### 5.2 Data Sets

Our data selection criteria are presented in Section 3.2. Here we present the libraries actually included in our study and describe reasons why we are unable to include all Rust crates in our data set.

At the start of our study, September, 2018, there were a total of 18,478 crates registered with the central Rust software repository. We eliminated all of the crates that could not compile or was no longer available. Note that Rust is under active development, and a particular crate may compile with one version of the compiler but not another. Afterwards, our data set contains 15,097 crates, which represent 81% of the total registered crates, and 97% of the registered crates that contain some Rust code and are syntactically correct.

To obtain a "popular" subset, we fetch the per-crate downloads numbers from crates.io and select the most downloaded crates that account for ninety percent of the downloads from crates.io. These 500 or so crates form a group we call *most downloaded*. From these most downloaded crates we were able to compile 473 crates.

To create a group of perhaps "higher quality" crates, we use the crates that comprise the application Servo, a web browser engine from Mozilla, as another group. Servo is implemented as a collection of approximately fifty crates and together with all its dependencies, it comprises almost four hundred crates. As Servo is created by many of the same developers actively developing the Rust language itself, we posit that it represents a more expertly developed piece of Rust software. Note that the crates in this group do overlap partially with the most downloaded and complete groups.

Finally, our fourth group contains the same crates as in the most downloaded group, but with the crate's contents as it existed in June 2019 on crates.io, approximately ten months after the first analysis.

### 5.3 Results

We provide answers the research questions from Section 3.1 based on the results of our experiments.

| Abstraction | crates.io (%) | Most Downloaded (%) | Servo (%) |
|---|---|---|---|
| Any | 29.4 | 52.5 | 54 |
| Blocks | 27.5 | 47.7 | 51.7 |
| *Declared Unsafe* Functions | 17.1 | 47.7 | 39.7 |
| Traits | 1.2 | 4.2 | 4.4 |
| Trait Implementations | 5.9 | 14.8 | 18.2 |

**Table 1: Percentage of Rust crates with *Unsafe Rust* based on abstraction type.**

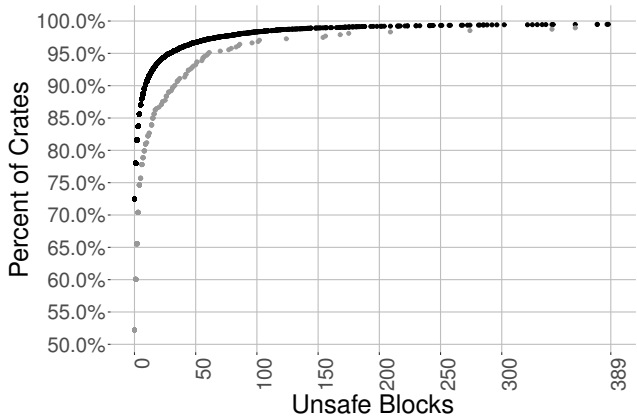

**Figure 3: RQ1: Cumulative Distribution of Unsafe Blocks**

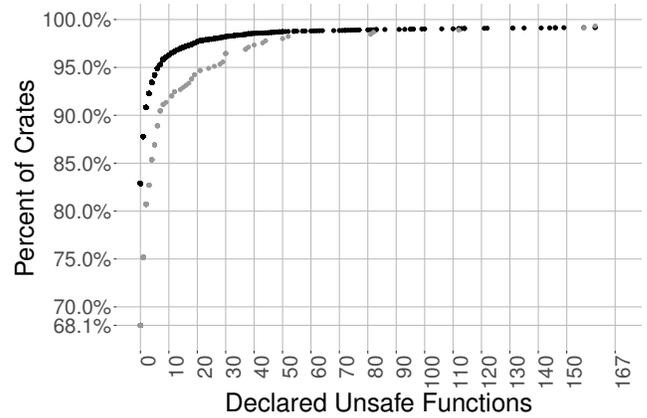

**Figure 4: RQ1: Cumulative Distribution of Declared Unsafe Functions**

**5.3.1 RQ1: How much do developers use *Unsafe Rust*?** Table 1 shows the percentages of the crates use **unsafe**, broken down by the type of abstraction. Overall, 29% of crates directly include some sort of *Unsafe Rust* in them. More popular crates are more likely to use **unsafe** as 52.5% of the most downloaded crates contain *Unsafe Rust*. Of these, only a few crates (about 15-17%) explicitly mark functions as **unsafe** (which then propagate the **unsafe** to other portions of code). The **unsafe** trait and trait implementations are used by only a relatively small number of crates. Compared to the larger ecosystem, the crates that comprise the Servo project are more likely to use *Unsafe Rust*.

*Blocks:* Figure 3 shows the cumulative distribution of unsafe blocks per crate, with crates.io set in black and the most downloaded crates in grey, and the maximum value capped at 99.5% of the crates for clarity. The long tail of the CDF (Cumulative Distribution Function) exists primarily because of autogenerated code, either from C to Rust translators, hardware description files, or "safe" Rust wrappers around C library functions.

The number of **unsafe** blocks per crate is small for the majority of the crates, more than 90% of the crates have fewer than ten **unsafe** blocks. The most downloaded crates use **unsafe** more often than all crates. This occurs because these crates are more likely to use **unsafe** to extract performance optimizations, and they often exist to help interface with existing C libraries.

*Functions:* Figure 4 shows the cumulative distribution of *declared unsafe* functions per crate for the crates.io set (in black) and most downloaded crates (in dark grey), capped at 99%. A very small number of crates have thousands of *declared unsafe* functions. These crates are typically low-level support crates for embedded devices or C library bindings.

As in the **unsafe** blocks case, the per-crate number of *declared unsafe* functions is small with 90% of the crates containing fewer than two *declared unsafe* functions.

While there is a compiler warning if a block is unsafe without using any unsafe operation, no such warning is generated for functions. One use of an unsafe function is to warn the library user that preconditions must be satisfied, otherwise memory corruption or a data-race may occur. We performed an additional analysis on the declared unsafe functions to determine if the function uses an unsafe operation. From all the crates with at least one declared unsafe function, 84% of all crates.io and 71% of most downloaded crates, all the declared unsafe functions are entirely *Safe Rust*. We identity two possible reasons for which a declared unsafe function executes no unsafe operations: (1) the library developer performed a careful analysis and determined that preconditions are necessary to prevent memory corruption and data races; or (2) the unsafe attribute was incorrectly placed.

*Traits:* Declaring an **unsafe** trait is rare in the Rust ecosystem, with only a little over one percent of crates making use the feature.

*Implementations:* As seen in Table 1, only six percent of all crates provide an implementation of an **unsafe** trait. Almost 40% of those are of two traits from the Rust standard library: Send (28%) and Sync (13%). These traits are fundamental for Rust's concurrency. They do not have any declared methods, and they are used as a declaration by the programmer that the objects implementing them are safe to change ownership to another thread (Send) or can be shared between threads (Sync).

**Summary:** *Unsafe Rust* is used in a little more than a quarter of all crates, but the number of explicit **unsafe** uses per crate is small for most crates. Despite the potential issues with using **unsafe**, the most downloaded crates are *more* likely to use **unsafe** than the average crate. In general, developers tend to avoid exposing the unsafety to other code by rarely marking functions as **unsafe** and typically avoiding **unsafe** traits. However, the crates for Servo use **unsafe** functions and **unsafe** traits two or three times as frequently as general crates, which is consistent with the growing preference in the Rust community that **unsafe** is encapsulated at a higher level than individual functions. That is, exposing **unsafe** functions is acceptable as long they are eventually enclosed within a safe interface.

| Analysis | crates.io (%) | Most Downloaded (%) |
|---|---|---|
| Conservative | 44.8 | 38.9 |
| Optimistic | 53.8 | 43.9 |

Table 2: Percentage of Crates With Only Safe Functions

|  | Dependencies With Unsafe (%) | No Unsafe in Dependencies (%) |
|---|---|---|
| Crate With Unsafe | 23 | 12 |
| No Unsafe in Crate | 38 | 27 |

Table 3: Unsafe in Crate and Dependencies

|  | crates.io (%) | Most Downloaded (%) | Servo (%) |
|---|---|---|---|
| Unsafe Function Call | 79.3 | 66.9 | 74.9 |
| Dereference C-Style Pointer | 16.2 | 19.5 | 21.5 |
| Global Variable | 3.7 | 11.1 | 0.21 |

Table 4: Unsafe Operations in Unsafe Blocks

|  | crates.io (%) | Most Downloaded (%) | Servo (%) |
|---|---|---|---|
| Unsafe Function Call | 89.2 | 64.0 | 78.2 |
| Dereference C-Style Pointer | 6.7 | 26.0 | 19.1 |
| Global Variable | 3.9 | 9.8 | 1.1 |

Table 5: Unsafe Operations in Unsafe Functions

#### 5.3.2 RQ2: How much of the Rust code is *Safe Rust*?
Table 2 presents the percentage of crates containing only *safe* functions, when the unsafe generated by compiler is ignored and the standard Rust library is considered entirely safe.

Figure 5 shows the cumulative distribution of the *possibly unsafe* functions for all crates, capped at 95%. The optimistic analysis is shown in light grey and the conservative analysis is shown in black. The difference between the results of the two analyses is about 10% for crates with a small number of *possibly unsafe* functions, and gets smaller than 1% for crates with tens of *possibly unsafe* functions.

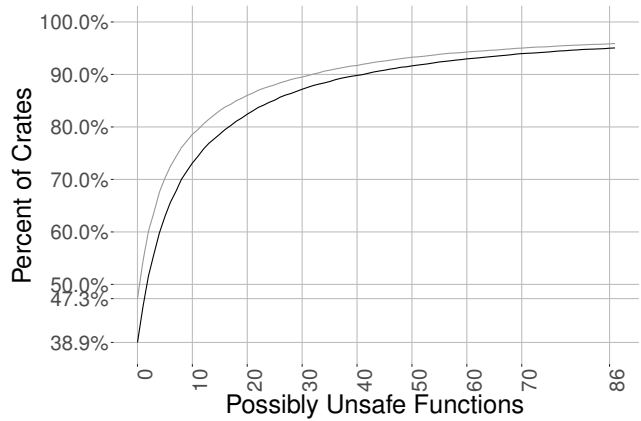

Figure 5: RQ2: Declared Safe and Possibly Unsafe Functions Distribution

To better understand the difference between the perceived safety (more than two thirds of the crates do not contain an **unsafe** abstraction, Table 1) and the static safety guarantees (less one third of crates are entirely *Safe Rust*) of Rust, we inspected crate dependencies. On average, across all crates in our dataset, a crate depends on twelve other crates. As shown in Table 3, only 27% of total crates contain no *Unsafe Rust* and only use dependencies which contain no *Unsafe Rust*. Importantly, 38% of crates include no **unsafe** in their own implementation, but rely on dependencies which do use **unsafe**.

**Summary:** While only less then one third of crates directly use **unsafe**, over half of crates include *Unsafe Rust* somewhere in the aggregate source code once dependencies are considered. This illustrates the difference between Rust's perceived safety and what is actually statically guaranteed. This also burdens developers trying to understand their software's exposure to *Unsafe Rust* as 38% of crates appear to avoid **unsafe**, yet contain it in their dependencies.

#### 5.3.3 RQ3: What Unsafe Rust operations are used?
Table 4 and Table 5 present the most frequent *Unsafe Rust* operations in **unsafe** blocks and *declared unsafe* functions, respectively. Only the *Unsafe Rust* operations present in more than 1% of crates are displayed.

Primarily, a function may be *declared unsafe* because it contains *Unsafe Rust* operations that are not enclosed in an **unsafe** block. Other reasons for declaring a function **unsafe** are: it is imposed by a **trait**, it has a precondition that must be satisfied, and it has a C-Style pointer argument that can be invalid. These cases are not directly using any *Unsafe Rust* operations, and thus were excluded from the analysis.

We observe that calls to *declared unsafe* functions are a majority of the *Unsafe Rust* operations used. The most downloaded crates and Servo use C-style pointers with greater frequency. One possible explanation is that these applications interface with C libraries for speed or because those libraries are not yet available in Rust.

**Summary:** Calls to **unsafe** functions are the majority of the *Unsafe Rust* operations. We need to understand if the functions are **unsafe** because they are implemented in C or they are Rust functions. This is investigated in the next research question.

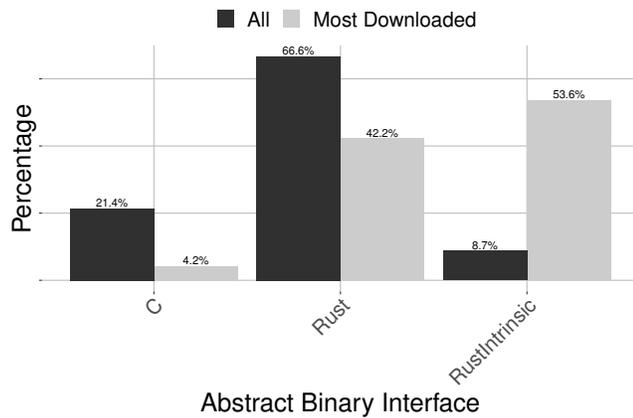

Figure 6: RQ4: Calls of Declared Unsafe Functions

| Abstraction | Same (%) | Increase (%) | Decrease (%) |
|---|---|---|---|
| Blocks | 82 | 10 | 8 |
| Functions | 87 | 7 | 6 |

Table 6: Unsafe Use in Most Downloaded Crates

#### 5.3.4 RQ4: What type of unsafe functions are called?
Figure 6 presents the distribution of the abstract binary interface of the declared unsafe functions called from both *declared unsafe* functions and `unsafe` blocks.

Among all crates from crates.io the most frequent calls are to Rust unsafe functions (65%), followed by calls to C functions (22.5%) and Rust intrinsics (special functions made available by the Rust compiler). Among the Rust unsafe calls, 47.6% are calls to the *Rust Core Library*. Of these `unsafe` functions in *Core*, 36.4% are of functions in the *ptr* module used to manually manage memory through C-style(raw) pointers and 40% are calls of functions that are `unsafe` wrappers to SIMD instructions and architecture-specific intrinsics. Thus, a significant source of *Unsafe Rust* is caused by the definition and use of C-style pointers, either by direct dereferencing or by calling `unsafe` functions from the *core::ptr* module that allow pointer arithmetic and access to the values stored at the pointer location.

The most downloaded crates use the Rust intrinsics much more frequently because intrisics provide operations used for I/O memory access and common atomic operations. The most downloaded crates include more libraries that extend Rust with I/O access and concurrency primitives. The very small number of C-style calls in most downloaded crates is explained by the fact that the most downloaded crates are more likely to contain crates that are wrappers for commonly used libraries. From all the calls to Rust unsafe functions, 47.2% are to function from the *Rust Core Library* of which 27% are to *core::ptr* functions. The wrappers to SIMD instructions are significantly used, as the calls to Rust intrinsics are much higher.
**Summary:** Calls to C functions are not the majority of the calls to `unsafe` functions. Implementing some C libraries in Rust, will remove some *Unsafe Rust* operations, but the majority of `unsafe` function calls are to Rust functions and Rust intrinsics. We conclude that the unsafe code is not encapsulated behind the public interface and the developers use the `unsafe` function when available.

#### 5.3.5 RQ5: Does the use of *Unsafe Rust* change over time?
To answer RQ5, we count the number of `unsafe` blocks and functions in the same set of most downloaded crates at two different time points: September 2018 and June 2019. Since the Rust compiler API changes frequently, we modified the Rust plugin to be able to compile updated crates. However, `unsafe` is counted in the same way, despite using two different versions of the tool.

A majority of crates (over 80%) contained the same number of `unsafe` blocks and functions in both versions. Of the crates that did change their `unsafe` usage, approximately half increased while the other half decreased.
**Summary:** We conclude that there are no significant trends in the use of `unsafe` over a period of ten months, with only a small increase in the use of `unsafe`.

### 5.4 RQ6: Why do Rust developers use unsafe?
To answer the RQ6 research question, we created a survey and posted it on the Rust Subreddit [1], and collected data from twenty respondents. The survey asked why they use `unsafe` and how they ensure correctness when using `unsafe`.

The first question asked Rust developers to select one or more reasons for why they use `unsafe` in their code. A majority (55%) indicated the use *Unsafe Rust* for higher performance, with *Safe Rust* being too restrictive as the next most common reason (40%). The other reasons selected include: the *Safe Rust* alternative is too verbose or complicated (25%), needed to make the code compile (10%), and faster to write code with *Unsafe Rust* (5%). Further, respondents provide other reasons, including: implementing fundamental data structures, custom concurrency primitives and system calls, interaction with specialized hardware, and integration with C or other languages.

To expand on the general reasons for using *Unsafe Rust*, the next question asked what operations the developers use that require the `unsafe` keyword. Of the respondents, 45% report using `unsafe` to call a non-syscall external C function, 25% to call an unsafe Rust function, 25% to work with C-style pointers, and 5% to work with SIMD intrinsics. No respondents selected to perform a system call or to access a static variable.

The final question asked the Rust developers for the steps they take to increase their confidence that their *Unsafe Rust* code is correct. Most respondents (65%) indicated they read the code very carefully, until they convince themselves that the code is correct. Another frequently used technique (55% selected this option) is adding runtime checks to prevent memory corruption. Half of the respondents write more unit tests for the function or method that uses unsafe Rust. Other steps developers take to increase confidence in the correctness of their `unsafe` code include: having discussions with experienced Rust developers in person or online, reading the documentation and Rust books, creating theoretical proofs, using available test generation, running fuzzing and analysis tools, and using Miri [17].

**Summary:** We conclude that the Rust developers use *Unsafe Rust* mostly because *Safe Rust* is too restrictive and to achieve better performance, but they are aware of the potential challenges when using `unsafe` and they use more care, test more thoroughly, and deploy tools and analysis to increase confidence in their code correctness.

## 6   Towards a Safer Rust

Our analysis shows that while publicly available Rust libraries rarely use the `unsafe` keyword (even very popular libraries), most of them are still not *Safe Rust*, because of `unsafe` use in dependencies.

For Rust users that require more robust and reliable checking of their code, future improvements to the Rust compiler or the associated tool infrastructure are required. A particularly ambitious mechanism would be to require an automatic correctness proof (in the style of Verifiable C [4]) for every use of `unsafe`. Jung *et al.* [18] used this approach to formally prove the correctness of a handful of Rust functions. The Rust functions were manually translated to a simplified version of the language, called $\lambda_{Rust}$ and using the Iris framework [19] the authors provided machine-checked proofs of correctness of several libraries and uncovered a previously unknown soundness bug in Rust. This approach is unlikely to scale to all Rust code, but may be feasible for the standard library or other commonly used libraries to help minimize the amount of code that is not checked for safety.

Another more feasible short-term approach is to address only `unsafe` function calls, which is the most common use of `unsafe`. Here, programmers could annotate function calls with suitable pre- and post-conditions they expect to hold when calling the function. Various static and dynamic checks then check the code to ensure the conditions are satisfied. This annotation would enable programmers to safely use functions that are intrinsically unsafe (i.e., configuring a hardware peripheral) or optimize performance with `unsafe`. However, this approach does add an additional burden for programmers and, possibly, run time overhead.

Another approach is to aid programmers in reasoning about the safety of any external code they use. Rust already includes a compiler directive that generates an error if `unsafe` is used, yet we discovered a library that included the directive but still used `unsafe` by overriding it at the function-level. Strengthening tools so programmers have automated checks on unsafe code in dependencies would significantly help programmers manually audit their codebase to better enable "pragmatic safety". For example, the Rust compiler should be able to identify functions in a given library that are *implicit safe*, and print a possible call chain that includes calls to *Unsafe Rust*. Implementing this in the compiler would make this tool easy to use, and it would remain up to date as the internal compiler API changes.

The development of the above mentioned tools requires a definition of what *Unsafe Rust* actually means and which behaviors are undefined. Despite Rust 1.0 being around for four years, the guidelines on how to write `unsafe` Rust code are still preliminary [35]. As this effort progresses, developing tools around *Unsafe Rust* will be more manageable and maintainable.

The Rust library ecosystem tool, crates.io, should also help alert developers to uses of `unsafe`. Anecdotally, over 100 unsound uses of `unsafe` in a popular Rust web framework, Actix [31], were only discovered and partially fixed by the Rust community after an online post by a concerned user [37]. Even a year later, the library still contains uses of `unsafe`, with soundness concerns still present. Motivated by this example, we propose the following changes to the crates.io interface: i) a new tag or badge for crates that include *Unsafe Rust*; ii) a dependency tree for each library with the crates that use *Unsafe Rust* clearly marked; and iii) a list of code reviews for any *Unsafe Rust*. Previous research established that code review in open source software communities is common and successful in eliminating a large number of errors [7, 26, 27]. Alami *et al.* [2] find that open source software developers develop a mature attitude to negative feedback and improve their code through a cycle of review, rejection, and improvement.

The implementation of our proposals will help the community inspect libraries and help guide new Rust developers on what *Unsafe Rust* code to trust.

## 7   Threats To Validity

The internal validity threats we identify are confounding factors and sampling bias. The *Unsafe Rust* segments we identify may be in unreachable or dead code that may not ever be executed for a particular library. Future work can use additional static analysis tools to identify uses of `unsafe` that reside in unreachable code. Sampling bias is also a possibility. We intended to be inclusive in the libraries we analyze, and we believe the way we segmented crates is makes our analysis representative of the larger Rust ecosystem. The crates included in our study are the ones that could be compiled on an Ubuntu operating system, and doe not include crates that need Windows and MacOS libraries.

Rust is under active development and it is possible that programming style and use of *Unsafe Rust* changes in time. We identify this as an external validity threat. The percentage of *safe* libraries may change when considering only "newer" libraries.

## 8   Related Work

We present throughout the paper related work relevant to the motivation and techniques used. In this section, we focus on related work that investigates how software engineers use various features in other programming languages.

**C/C++ language features:** Eyolfson *et al.* analyze the use of `const` annotations in seven medium and large open source C++ projects and find that software developers use correct annotations in most cases, missing immutability labels in only 6% of the unannotated methods [13]. In Rust, the variables are immutable by default, and the programmers opt-in by labeling a variable as mutable. This study suggests that C++ programmers are using language features to write software with fewer errors. In our study, we observe the same tendency as the majority of crates are free of direct uses of *Unsafe Rust*.

Casalnuovo *et al.* [9] studied many C/C++ projects to measure the programmers' use of `assert`. They find that a majority of the projects use more than a minimal number of `assert` statements.

The use increases with the length of time the developer is directly involved with that function. If Rust developers behave similarly, then over time the number of assertion statements in Rust code should increase, helping to protect *Unsafe Rust* operations. However, developers may instead view assertions as not needed in Rust.

Undefined behavior can be particularly problematic in code as the compiler can correctly generate machine code that does not match the programmer's expectations. As with the case of using undefined integer overflow to check a buffer's length [12], undefined behavior can lead to security vulnerabilities. Wang *et al.* [38] implement a tool to detect undefined behavior based on differences resulting from the compiler optimization level used. They find that 40% of the 8,575 Debian Wheezy packages that contain C/C++ code exhibit this behavior, and identify 160 real bugs in production open source code.

Further, understanding undefined behavior can be difficult, as Memarian *et al.* [23] demonstrate by surveying over 200 experienced C developers and asking 85 difficult questions about the semantics of C. The researchers conclude that in many instances there was no agreement among the participants on the actual code behavior. As *Unsafe Rust* can easily introduce undefined behavior into Rust code, these case studies suggest *Unsafe Rust* should be used very cautiously.

**Java**: Java is a safe language, but the runtime provides a "backdoor" that permits the circumvention of Java's safety guarantees to enable high-performance systems-level code. Mastrangelo *et al.* [22] perform a large-scale analysis of Java bytecode to determine how these unsafe capabilities are used in real world. The authors determine that 25% of the Java code analyzed depends on unsafe Java code. One explanation for why the 25% of the analyzed Java code lacks safety because of use of unsafe API may be that it is not an integral part of the language, like in Rust, and it is not exposed by the java standard libraries. Huang *et al.* [16] study unsafe crash patterns and implement a bytecode-level transformation that introduces runtime checks to help diagnose and prevent some memory errors caused by the use of the unsafe API.

**Swift:** Swift, introduced by Apple, is intended to replace Objective-C, introduced a new error handling mechanism using exceptions that is not present in Objective-C. Cassee *et al.* [10] execute a large-scale study to identify if Swift developers switched to the new mechanism. They find that about half of the projects do not use the new error handling mechanism. Of the projects that do, some follow some guidelines, but most do not follow the more complex error-handling recommendations. This fallback to known patterns may also exist in Rust systems code, as many Rust systems developers likely have experience writing C. That is, Rust developers may use *Unsafe Rust* to enable using the C-style code patterns they are familiar with. Further studies are required to understand if this is the case.

## 9 Conclusions

In this research, we explore whether Rust is, in fact, being used to ensure memory safety and avoid concurrency bugs. To answer a number of research questions, we develop a technique to construct an extended call graph for Rust and an analysis that determines whether functions possibly include *Unsafe Rust* or not, depending on their dependencies. We conduct a number of experiments on Rust code, using the results to answer our research questions. Across our dataset, we find that a majority of crates include functions which are possibly unsafe. We also find that unsafe function calls are the most common use of unsafeness, and the unsafeness is through library dependencies rather than through the use of the keyword `unsafe`. Perhaps nonintuitively, we find that the most downloaded crates have more unsafe code than other crates. From these results, it is difficult for users to know if their code is safe, and thus we present recommendations for helping users understand when they are using Unsafe Rust in their software.